\documentclass[aps, prl, twocolumn, showpacs, amsmath, amssymb, superscriptaddress, reprint, longbibliography]{revtex4-1}
\usepackage[utf8]{inputenc}
\usepackage[T1]{fontenc}
\usepackage{xcolor,amsmath,amssymb}

\usepackage{graphicx}
\usepackage[textsize=footnotesize]{todonotes}

\definecolor{darkblue}{rgb}{0,0,0.6}
\definecolor{darkred}{rgb}{0.6,0,0}
\usepackage[colorlinks=true,urlcolor=darkblue,citecolor=darkblue,linkcolor=darkred,hyperfootnotes=false]{hyperref}

\newcommand{\iden}{{\bf 1}}
\newcommand{\ind}[1]{_{\mathrm{#1}}}

\newcommand{\kk}{\boldsymbol{k}}

\newcommand{\rr}{\boldsymbol{r}}

\newcommand{\eepsilon}{\boldsymbol{\epsilon}}

\newcommand{\ssigma}{\boldsymbol{\sigma}}

\DeclareMathOperator{\tr}{tr}

\begin{document}

\title{Fault orientation in damage failure under compression}

\author{Véronique Dansereau}
\affiliation{Univ. Grenoble Alpes, CNRS, ISTerre, 38000 Grenoble, France}

\author{Vincent Démery}
\affiliation{Gulliver, CNRS, ESPCI Paris, PSL Research University, 10 rue Vauquelin, 75005 Paris, France}
\affiliation{Univ Lyon, ENS de Lyon, Univ Claude Bernard Lyon 1, CNRS, Laboratoire de Physique, F-69342 Lyon, France}

\author{Estelle Berthier}
\affiliation{Institut Jean Le Rond d’Alembert, UMR 7190, CNRS and Université Pierre et Marie Curie, 75005 Paris, France}
\affiliation{Department of Physics, North Carolina State University, Raleigh, North Carolina, USA}

\author{Jérôme Weiss}
\affiliation{Univ. Grenoble Alpes, CNRS, ISTerre, 38000 Grenoble, France}

\author{Laurent Ponson}
\affiliation{Institut Jean Le Rond d’Alembert, UMR 7190, CNRS and Université Pierre et Marie Curie, 75005 Paris, France}


\begin{abstract}
The Mohr-Coulomb criterion is widely used in geosciences to relate the state of stress at failure to the observed orientation of the resulting faults.
This relation is based on the assumption that the fault occurs along a plane that maximizes the Coulomb stress.
Here, we test this hypothesis using an elastic, progressive damage model that implements the Mohr-Coulomb criterion at the local scale.
We find that the orientation of the fault is not given by the Mohr-Coulomb criterion. Instead, for minimal disorder, it corresponds to the most unstable mode of damage in the model, which we determine through a linear stability analysis of the homogeneously damaged state.
Our simulations show that microstructural disorder significantly affects the orientation of the fault, which, however, remains always far from the Mohr-Coulomb prediction.
\end{abstract}

\maketitle

In 1773, Charles-Augustin de Coulomb proposed his celebrated failure criterion for materials loaded under shear or compression~\cite{Coulomb1773}. 
He postulated that failure occurs along a fault plane when the applied shear stress $\tau$ acting on that plane overcomes a resistance consisting of two parts of different nature: a cohesion $\tau_c$, which can be interpreted as an intrinsic shear strength of the material, and a resistance proportional to the normal pressure, $\sigma_N$; this results in the Mohr-Coulomb (MC) failure criterion,
\begin{equation}\label{eq:MC_crit}
\left| \tau \right| = \tau_c + \mu \sigma_N. 
\end{equation}
Following the former work of Amontons~\cite{Amontons1699}, this dependence upon pressure led Coulomb to call it \textit{friction}, with $\mu$, the corresponding friction coefficient and $\phi = \tan^{-1}(\mu)$, the angle of internal friction. 
As a consequence, faulting should occur along the plane that maximizes the Coulomb's stress $|\tau| - \mu \sigma_N$. 
Its orientation with respect to the maximum principal compressive stress is given by the MC angle
\begin{equation}\label{eq:MC_angle}
\theta\ind{MC} = \frac{\pi}{4}-\frac{\phi}{2}.
\end{equation}
This work led to the so-called Anderson theory of faulting~\cite{Anderson1905}, which is widely used in geophysics to interpret the orientation of conjugate faults~\cite{Schulson2004} and the orientation of faults with respect to tectonic forces~\cite{Reches1987}. 
In this theory, $\theta\ind{MC}$ is uniquely a function of the internal friction angle $\phi$ and hence is independant of confinement and dilatancy.

At geophysical scales, some of the most destructive natural hazards, such as earthquakes and landslides, result from the sudden release of stored elastic energy through a failure process. 
Coulomb's theory of failure prevails to this day as the generic framework of interpretation and modelling of these ruptures in Solid Earth (e.g., for interpreting the changes in static Coulomb stress following earthquakes that trigger aftershocks~\cite{Stein1999}). 
However, important issues remain to be addressed regarding the applicability of this theory to compressive failure. If laboratory experiments on rocks~\cite{Byerlee1967, JaegerCook1979} and ice~\cite{Schulson2006, WeissSchulson2009} have successfully reproduced the MC failure envelope, the problem of the fault orientation has somewhat been overlooked~\cite{Haimson2010}.
Besides, while Coulomb's theory provides a simple instantaneous criterion for failure, it says nothing about the failure process itself. It is now widely accepted that the compressive failure of quasi-brittle materials does not occur suddenly, but instead involves the progressive nucleation of microcracks, which interact and finally coalesce to form a macroscopic fault~\cite{Lockner1991, Fortin2009, Renard2017}. 
It is not at all clear if this phenomenology is compatible with the MC theory, nor with the assumption that the fault orientation is given by the MC criterion. 

\begin{figure}
\begin{center}
\includegraphics[angle = 0, scale=0.57]{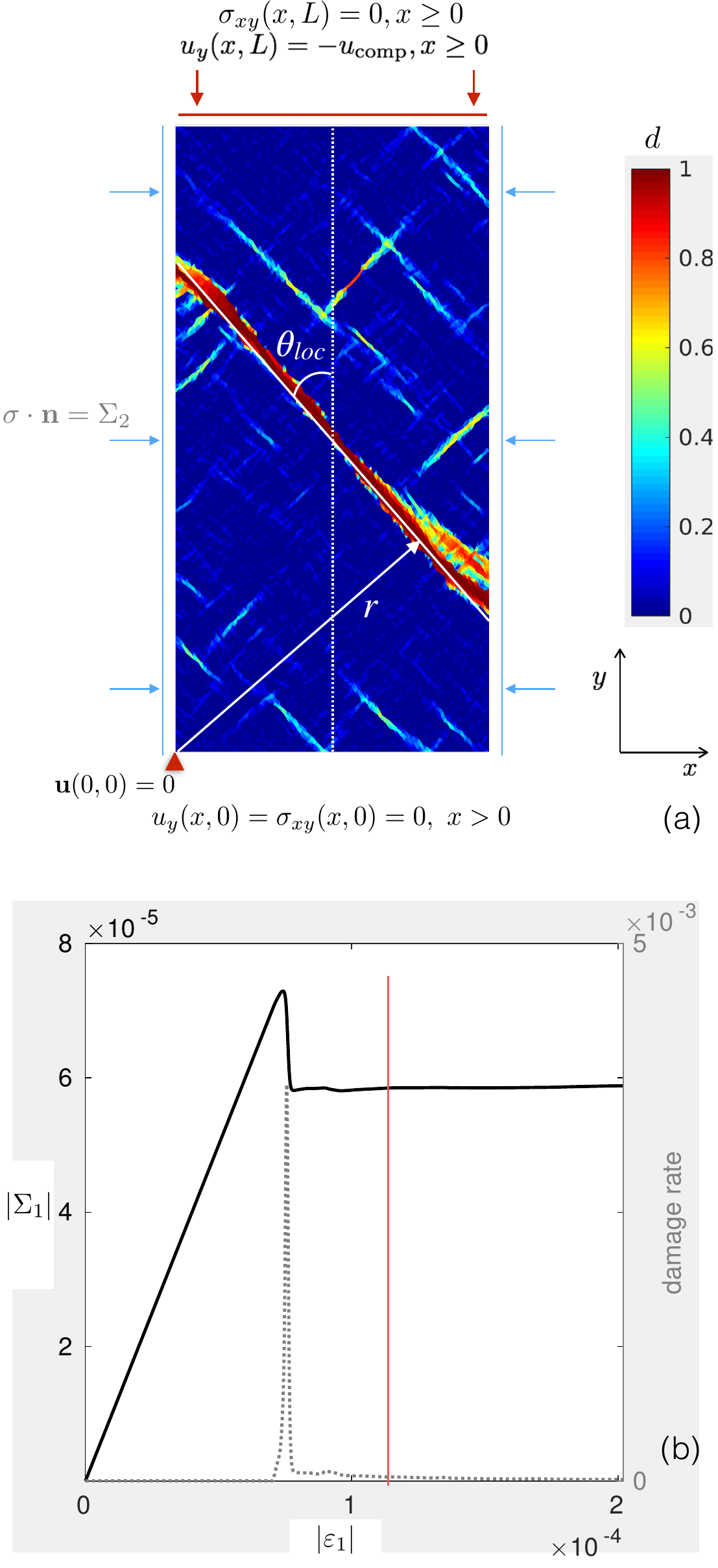}
\end{center}
\caption{(a) Domain and boundary conditions for the compression simulations and field of the instantaneous level of damage after the formation of a macroscopic fault in such a simulation. The solid white line and white arrow represent the location, $r$, and orientation, clockwise or anti-clockwise from the vertical, $\theta\ind{loc}$, of the fault as estimated by the projection histogram method. (b) Macroscopic stress $\Sigma_1$, vs. macroscopic strain $\varepsilon_1$, in a compression simulation using $R = 0$ and $\nu = 0.3$ with the red line indicating the timing of the instantaneous field of damage shown in (a); the dashed line indicates the damage rate.
}
\label{fig:stress_strain_damage}
\end{figure}

This progression towards macroscopic failure is well captured by continuum damage models, wherein microcrack density at the mesoscopic scale is represented by a damage variable and is coupled to the elastic modulus of the material~\cite{Zapperi1997, Lyakhovsky1997, Amitrano1999, Girard2010} (Fig.~\ref{fig:stress_strain_damage}).
In these models, a failure criterion is implemented at the local scale, that is, usually, the scale of the grid element. When the state of stress over a given element exceeds this criterion, the level of damage of this element increases, thereby decreasing its elastic modulus. 
Long-range elastic interactions arise from the stress redistribution initiated by the local drop in the elastic modulus.
It can induce damage growth in neighbouring elements and eventually trigger avalanches of damaging events over longer distances.

Such models have been shown to reproduce many features of brittle compressive failure, such as the clustering of rupture events along faults, as well as the power law distribution of accoustic events sizes prior to the macro-rupture observed in rock-like materials \citep{Tang1997, Zapperi1997, Girard2010, Amitrano2003}. 
They are thus relevant tools to study fracture and, in particular, the dependence of the angle of localization of damage on the parameters involved in damage criteria.

Here, we focus on the relation between the orientation of the macroscopic fault and the internal friction angle, $\phi$, involved in the MC theory, using a continuum, finite element damage model that implements the MC criterion at the local scale.
We show that the orientation of the simulated faults is not given by the MC angle (Eq.~(\ref{eq:MC_angle})).
Instead, we find that the most unstable mode of damage growth, computed from a linear stability analysis of the damage model, provides a good estimation of the fault orientation in case of minimal disorder.
Finally, we show that the orientation of the fault is sensitive to disorder.

Similar to \cite{Amitrano1999} and others, the model is based on a linear-elastic constitutive law and an isotropic, progressive damage mechanism for the elastic modulus, $E$, that involves a scalar damage variable, $d$. Here the level of damage is defined to be $0$ for an undamaged and $1$ for a ``completely damaged'' model element, and
\begin{equation}\label{eq:E_d}
E(d) = (1-d) E^0 ,
\end{equation}
with $E^0$ the Young's modulus of an undamaged element. 

In the numerical simulations, a two-dimensional rectangular sample of an elasto-damageable material with dimensions $L \times L/2$ is compressed with a stress $\Sigma_1$ by prescribing a constant velocity on its upper short edge with the opposite edge fixed in the direction of the forcing (Fig. \ref{fig:stress_strain_damage}a).
Plane stresses are assumed. 
A confining stress, $\Sigma_2$, can be applied on the lateral sides; in this case the confinement ratio $R=\Sigma_2/\Sigma_1$ is kept constant throughout the simulation.
Both the upper edge velocity and lateral confinement are small enough to ensure a low driving rate and small deformations. 
In this case, the model is described by the force balance and Hooke's law:
\begin{align}
\nabla \cdot \mathbf{\ssigma} & = 0 \label{mom_eqn}, \\
\ssigma & = \frac{E}{1+\nu}\eepsilon + \frac{E\nu}{1-\nu^2}\tr(\eepsilon)\iden,  \label{const_eqn}
\end{align}
with $\ssigma$ the stress tensor, $\eepsilon$ the strain tensor, and $\nu$ Poisson's ratio.

Equations \eqref{mom_eqn} and \eqref{const_eqn} are solved using variational methods on an amorphous grid made of more than $33000$ triangular elements. 
The model and numerical scheme are further detailed in \citep{long}. The simulated macrocospic stresses at failure reproduce a Mohr-Coulomb failure enveloppe with virtually the same value of $\mu$ as that prescribed at the micro scale \citep{long}, i.e. $\mu$ appears as a scale-independent property, in agreement with observations~\citep{WeissSchulson2009}, and the post-macro failure damage field shows a localized linear structure (Fig.~\ref{fig:stress_strain_damage}). 
A projection histogram calculation is used to determine its orientation, hereinafter referred to as the localization angle, $\theta\ind{loc}$~\citep{long}. 

\begin{figure*}
\begin{center}
\includegraphics[angle = 0, scale=0.40]{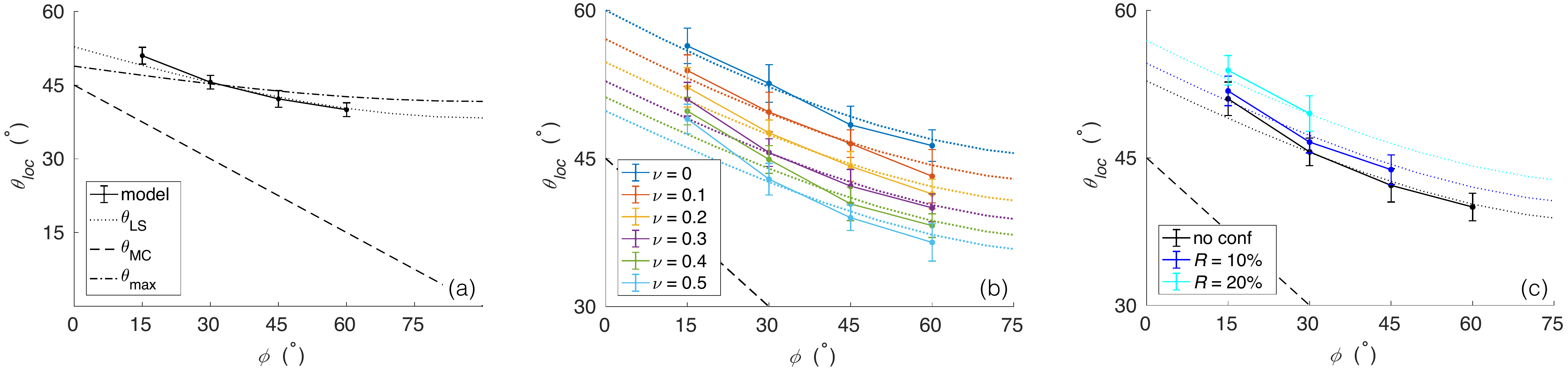}
\end{center}
\caption{(a) Mean $\theta\ind{loc}$ as a function of $\phi$ for an ensemble of 25 simulations with minimal disorder using identical boundary and loading conditions. No confinement is applied and $\nu = 0$. The black dashed line shows to the MC prediction, $\theta\ind{MC}$, the dotted line, the angle of the most unstable mode, $\theta\ind{LS}$ and the dashed-dotted line, the angle of maximal MC stress redistribution, $\theta\ind{max}$. The error bars represent $\pm 1$ standard deviation from the mean.
Mean $\theta\ind{loc}$ for (a) different values of Poisson's ratio without confinement and (b) different values of confinement ratio for $\nu = 0.3$.
}
\label{fig:angle_single} 
\end{figure*}

A first set of compression simulations representing a minimum disorder scenario was initialized with a field of cohesion that was uniform for all except one element chosen at random. For this inclusion, $\tau_c$ was initially 5\% weaker and was reset to the uniform value of its neighbors after its first damage event. 
Fig. \ref{fig:angle_single}a shows the mean estimated localization angle as a function of the internal friction angle, $\phi$, and Fig. \ref{fig:angle_single}b and \ref{fig:angle_single}c, the same results for different Poisson's and confinement ratios, respectively. 
Neither the value nor the variations of $\theta\ind{loc}$ with $\phi$ agree with the MC prediction.
In particular, the simulated fault orientation varies with Poisson's ratio as well as with confinement, a dependance that is not accounted for in the MC theory.

In order to understand how the localization of damage arises in these numerial simulations, we performed a linear stability analysis of the homogeneously damaged solution of the model. Damage evolution in our case can be written generally as
\begin{equation}\label{eq:damage_growth}
\frac{\partial d}{\partial t}(\rr,t)=F[\ssigma_0,d(\rr,t)].
\end{equation}
Note that the function $F$ is non-local: the evolution of damage at one point in the sample depends on the damage level everywhere in the specimen.
The linear stability analysis amounts to linearizing this evolution equation for a weakly heterogeneous damage field, $\delta d(\rr,t)= d(\rr,t)-d_0\ll 1$.
Assuming an infinite system, the problem is translation invariant and the linearization can be written as a convolution product of the damage field with the \textit{elastic kernel} $\Psi_{\ssigma_0,d_0}$~\cite{Weiss2014}:
\begin{equation}\label{eq:lin_stab}
F[\ssigma_0,d(\rr,t)]\simeq F[\ssigma_0,d_0]+\Psi_{\ssigma_0,d_0}*\delta d(\rr,t).
\end{equation}
The kernel $\Psi$ is reminiscent of the Eshelby solution for the mechanical field around a soft inclusion embedded in an infinite 2D elastic medium, which shows the same scaling behavior in $r^{-2}$~\cite{Eshelby1957}.

The elastic kernel is more easily calculated in Fourier space~\cite{long}. Due to the absence of characteristic length in the problem, the kernel does not depend on the magnitude of the wavevector, but only on its polar angle, $\omega$:
\begin{equation}\label{eq:kernel}
\tilde \Psi(\omega)=A\left[\sin(\omega)^2-\frac{1+\sin(\phi)}{2} \right]\times \left[\delta-\sin(\omega)^2\right],
\end{equation}
where $A=2 \Sigma_1 (1-\nu)(1-R)/(1-d_0)$ and
\begin{equation}
\delta=\frac{\nu-R}{(1+\nu)(1-R)}.
\end{equation}

The evolution of the damage field perturbations are inferred from Eqs.~(\ref{eq:damage_growth}, \ref{eq:lin_stab}).
The growth rate of harmonic modes $\delta d(\rr)\propto\cos(\kk\cdot\rr)$ is set by $\tilde\Psi(\omega)$, since the convolution product is a simple product in Fourier space.
The kernel is maximal and positive for $\sin(\omega^*)^2=[1+\sin(\phi)+2\delta]/4$. This means that (i) a homogeneous damage field is unstable and (ii) 
since the value of the kernel does not depend on the magnitude of the Fourier wavevector, all the wavevectors with the orientation $\omega^*$ diverge at the same rate. 
Hence, any linear combination of the Fourier modes with these wavevectors also diverges at the same rate and corresponds to a localization pattern with orientation $\pi/2\pm\omega^*$ (perpendicular to $\pm\omega^*$), i.e.,
\begin{equation}\label{eq:loc_angle}
\theta\ind{LS}= \arccos\left(\frac{\sqrt{1+\sin(\phi)+2\delta}}{2}\right)
\end{equation}
with respect to the direction of maximum principal compressive stress; we kept only the solution lying in $[0,\pi/2]$.
We expect this angle to correspond to the angle $\theta\ind{loc}$ measured in the simulations.
%
%
This prediction is in very good agreement with the results of the minimal disorder numerical simulations (Fig.~\ref{fig:angle_single}) and reproduces the observed dependence on Poisson's ratio and confinement.

Alternatively, the fault orientation is sometimes assumed to correspond to the angle where the stress redistribution is maximal after a damage event~\cite{Nicolas2017}.
This angle, $\theta\ind{max}=\arccos(\sqrt{[3+\sin(\phi)+2\delta]/8})$, determined from the elastic kernel (Eq.~(\ref{eq:kernel})) written in real space~\cite{long}, is different from the orientation of the most unstable mode, $\theta\ind{LS}$.
The difference arises because the localization mode emerges from the redistribution of damage in \textit{all} directions, which is not symmetric around $\theta\ind{max}$. As shown in Fig.~\ref{fig:angle_single}a, $\theta\ind{LS}$ clearly provides a better agreement with the simulations in the case of a single, evanescent defect.

However, real and, especially, natural materials are disordered and comprise many defects or impurities that can serve as local stress concentrators, initiating microcracking and leading to a regime of diffuse damage growth prior localization~\cite{Lockner1991}. 
To determine if and how this regime affects the final orientation of the macroscopic fault, we introduce disorder by drawing randomly the (adimensional) value of cohesion, $\tau_c$, of a proportion $a$ of the model elements from the uniform distribution $[1-\epsilon, 1+\epsilon]$, with the cohesion of the remaining proportion $1-a$ of the elements set to 1.

\begin{figure}
\begin{center}
\includegraphics[angle = 0, scale=0.62]{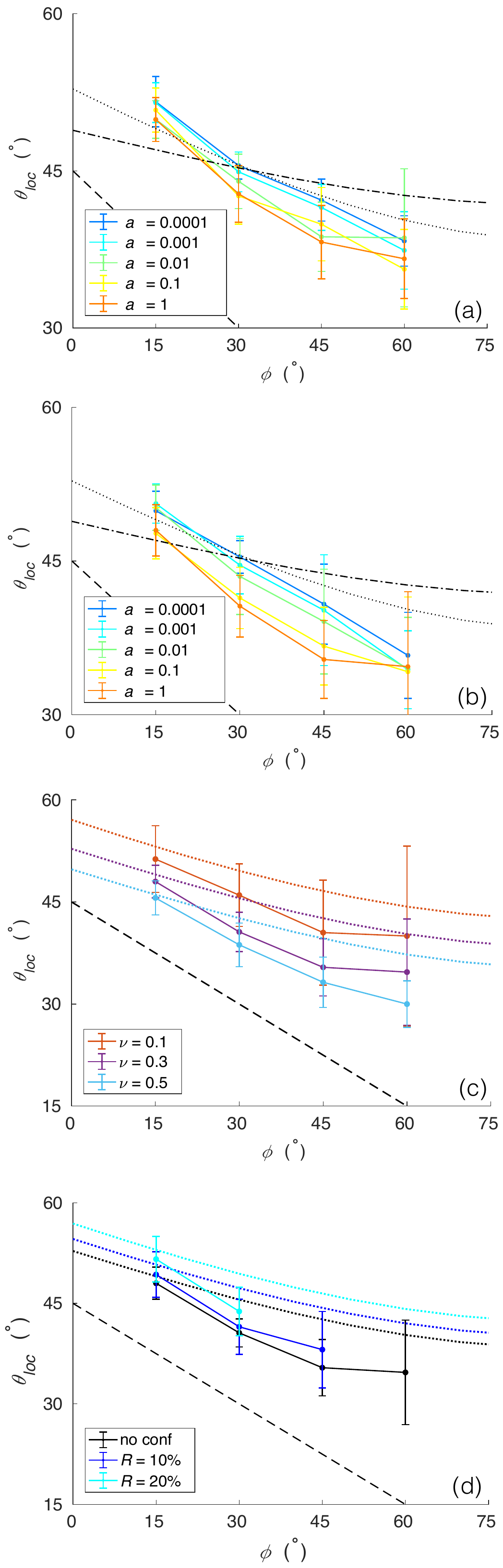}
\end{center}
\caption{Mean $\theta\ind{loc}$ as a function of $\phi$ for (a) weak disorder, $\epsilon=0.05$ and (b) strong disorder, $\epsilon=0.5$ and different values of $a$. No confinement is applied and $\nu=0.3$. Mean $\theta\ind{loc}$ for $a = 1$ and $\epsilon = 0.50$ (strong disorder) and (c) different values of $\nu$ without confinement and (d) different confinement ratios for $\nu = 0.3$. The black dashed line shows $\theta\ind{MC}$, the dotted lines, $\theta\ind{LS}$ and the dashed-dotted line, $\theta\ind{max}$.
}
\label{fig:disorder}
\end{figure}

We consider cases of weak ($\epsilon=0.05$, Fig. \ref{fig:disorder}a) and strong ($\epsilon=0.5$, Fig. \ref{fig:disorder}b) quenched disorder. 
In both cases, the value of $a$ is varied between $10^{-4}$, corresponding to a few ($\sim 3$) inclusions in a homogeneous matrix, and $a = 1$, for which all elements have a different critical strength.
Consistently with the minimum disorder case investigated above, the agreement with the prediction from the linear stability analysis is best for $a = 10^{-4}$ (Fig. \ref{fig:disorder}a, b). 
The deviation from the prediction increases with both the density of inclusions (i.e., $a$) and with the strength of the disorder (i.e., $\epsilon$), indicating that disorder significantly affects the fault orientation $\theta\ind{loc}$. 
In all cases however, $\theta\ind{loc}$ remains far from $\theta\ind{MC}$. Notably, a clear dependence on Poisson's ratio and on confinement is still observed (Fig. \ref{fig:disorder}c, d). This questions the estimation of internal friction or of applied stresses from faults orientation in natural settings~\citep{Anderson1905, Schulson2004, Reches1987}. 

Predicting the localization angle in a strongly disordered material remains an open question. A possible extension to this work would be to push further the analysis of the linear stability analysis. Modes around the most unstable mode may play a role in setting the orientation of the fault, which is particularly likely in the disordered case.

The discrepancy between the fault angle and the Mohr-Coulomb prediction indicates that compressive failure, even when it is not preceded by an extended regime of stable damage growth, results from the {\it collective} spreading of damage within the specimen. 
As such, the fault angle observed in our simulations was successfully captured from a stability analysis performed at the specimen scale. 
The role of elasticity, which redistributes the stress after a damage event and is responsible for interactions between microcrocaks, reflects in the dependence of the localization angle on the Poisson's ratio.
The fact that the criterion derived from the stability of a {\it single element} fails to predict the fault angle suggests that other local approaches that do not account for the interactions between damage events, e.g.,~\cite{Rudnicki1975}, may not describe damage localization accurately.
The critical comparison of both points of view is currently under progress.

\begin{acknowledgments}
V. Dansereau is supported financially by TOTAL S.A.. E. B. has been supported by the program Emergence from UPMC.
We thank D. Kondo for usefull discussions.
\end{acknowledgments}



%

\end{document}